# Generalized BF Theory in Superspace as Underlying Theory of 11D Supergravity

Hitoshi NISHINO[1] and Subhash RAJPOOT[2]

*Department of Physics & Astronomy*
*California State University*
*Long Beach, CA 90840*

**Abstract**

We construct a generalized $BF$ theory in superspace that can embed eleven-dimensional supergravity theory. Our topological $BF$ theory can accommodate all the necessary Bianchi identities for teleparallel superspace supergravity in eleven-dimensions, as the simplest but nontrivial solutions to superfield equations for our superspace action. This indicates that our theory may have solutions other than eleven-dimensional supergravity, accommodating generalized theories of eleven-dimensional supergravity. Therefore our topological theory can be a good candidate for the low energy limit of M-theory, as an underlying fundamental theory providing a 'missing link' between eleven-dimensional supergravity and M-theory.

---

[1]E-Mail: hnishino@csulb.edu
[2]E-Mail: rajpoot@csulb.edu

## 1. Introduction

It has been well-known that Chern-Simons theories in three-dimensions (3D) or topological $BF$ field theories [1] have important physical significance in physics. For example, 3D gravity theory is nothing other than a Chern-Simons theory in 3D, which is exactly soluble [2]. On the other hand, there have been considerable developments associated with M-theory [3] as the most fundamental theory unifying most of the known superstring theories [4] in 10D, as well as lower-dimensional strings, whose low energy limit is supposed to be described by 11D supergravity theory [5].

Considering such developments in M-theory [3], a common expectation nowadays seems that there must exist a more explicit formulation in 11D as a generalization of 11D supergravity [5], based on topological formulation, or on more enlarged gauge groups, dualities or higher-derivative $R^2$-theories. For example, Chern-Simons supergravity formulation [6][7][8] is one of such trials for exploring an underlying theory of M-theory [3]. For example, in our recent paper [9], it has been shown that an action for fundamental extended objects for Chern-Simons supergravity [6][7][8] in 11D for the group $OSp(32|1)$ coincide with Type II Green-Schwarz superstring action [4] in a certain limit. Another example is the reformulation of 11D supergravity [5] as a first-order topological field theory with certain constraints [10]. This formulation yields the action proposed by D'Auria and Fré some time ago [11]. As for searching for the fundamental gauge group of 11D supergravity [5], it is argued in ref. [12] that the symmetry of M-theory [3] is to be the group $OSp(64|1)$, and is further developed in [13] that 11D supergravity [5] is described by a non-linear realization based on the group $E_{11}$, and the gravitational degrees of freedom can be described by two fields related by duality. In ref. [14], a MacDowell-Mansouri $R^2$-type action for the superalgebra $OSp(8|1)$ was proposed as possible low-energy limit of M-theory [3]. Even though a particular link with M-theory [3] or 10D superstring [4] was shown in a certain limit in Chern-Simons supergravity [9], or some enlarged gauge group formulations [12][13] provide some scenarios, there still seems to be some gap between these formulation and M-theory [3], in particular 11D supergravity in superspace [15]. Moreover, the drawback of the topological formulation in [10] is the usage of component language which is not always convenient to control certain generalization of 11D supergravity [5][15], such as M-theory corrections, compared with superspace formulation [16]. For such a superspace formulation of M-theory, there is a recent trial [17], of construction of eleven-dimensional superspace with superspace coordinates and a finite M-theory using non-anticommutative geometry. However, the most crucial supersymmetric invariance of the action in [17] remains to be confirmed. The need of superspace formulation may be attributed to the most common expectation that a desirable fundamental underlying theory must have a very simple appearance possibly formulated in superspace, with topological/geometrical features, but at the same time, it should be rich and intricate enough to accommodate such complicated theory as 11D supergravity [5].

In our present paper, we present a simple topological 'generalized' $BF$ theory formulated in superspace [18] that can possibly unify the two important recent theories, *i.e.,* topological $BF$ or Chern-Simons theory [1] and M-theory [3] in terms of superspace language [18]. We present a simple superspace lagrangian, in which all the necessary superspace Bianchi identities in 11D [15] are 'automatically' accommodated, as solutions of the superfield equations obtained from our superspace action. To this end, we also utilize the recent result in teleparallel superspace formulation [19], in which the local Lorentz symmetry in the conventional 11D superspace [15] is no longer manifest. This is because in teleparallel superspace, we do not need to introduce the supercurvature $R_{ABc}{}^d$ which had been the main obstruction of embedding certain Bianchi identity components in topological superfield equations [19]. Since our formulation is based on superspace, all the usual features of space-time supersymmetry are built-in from the outset, in sharp contrast with other formulations such as Chern-Simons supergravity [6][7][8][9] in which space-time supersymmetry is not manifest.

## 2. Review of Teleparallel Superspace

Before presenting our lagrangian, we briefly review the important ingredients in our teleparallel superspace [19] for 11D supergravity [15] which plays a central role in our formulation. In the teleparallel superspace [19], we have no manifest local Lorentz covariance, and therefore these is no need of supercurvature $R_{ABc}{}^d$, in contrast to the conventional superspace formulation [15][18]. As will be seen, the absence of supercurvature in such superspace [19] enables us to consider the embedding of all the necessary Bianchi identities into the superfield equations in our topological superfield theory. As such, we have only two basic geometrical superfield strengths, *i.e.,* the anholonomy coefficients which is equivalent to a torsion superfield denoted by $C_{AB}{}^C$ and a 4-th rank antisymmetric superfield strength $F_{ABCD}$ which is similar to the conventional 11D superspace [15][18].

The Bianchi identities in our teleparallel superspace supergravity are [19]

$$\tfrac{1}{2} E_{[A} C_{BC)}{}^D - \tfrac{1}{2} C_{[AB|}{}^E C_{E|C)}{}^D \equiv 0 \ , \tag{2.1a}$$

$$\tfrac{1}{24} E_{[A} F_{BCDE)} - \tfrac{1}{12} C_{[AB|}{}^F F_{F|CDE)} \equiv 0 \ , \tag{2.1b}$$

where the derivative $E_A$ has no Lorentz connection. It replaces the Lorentz covariant derivative $\nabla_A$ in the conventional superspace formulation [18], and all the torsion superfields $T_{AB}{}^C$ in the conventional superspace [18] are replaced by the anholonomy coefficients $C_{AB}{}^C$. To be more specific,

$$E_A \equiv E_A{}^M \partial_M \ , \quad C_{AB}{}^C \equiv (E_{[A} E_{B)}{}^M) E_M{}^C \ , \quad [E_A, E_B\} \equiv C_{AB}{}^C E_C \ . \tag{2.2}$$

The supercurvature term in (2.1a) is now absent, due to the lack of manifest local Lorentz covariance. Needless to say, there is no Bianchi identity for the supercurvature, such as $\nabla_{[A} R_{BC)d}{}^e - \cdots \equiv 0$, either. As usual, the $\theta = 0$ sector of $F_{abcd}$ corresponds to the 4-th rank component field strength in 11D supergravity [5][15].

More explicitly, the constraints at the mass dimensions $1/2 \le d \le 1$ for our teleparallel superspace [19] are

$$C_{\alpha\beta}{}^c = +i(\gamma^c)_{\alpha\beta} \ , \qquad F_{\alpha\beta cd} = +\tfrac{1}{2}(\gamma_{cd})_{\alpha\beta} \ , \tag{2.3b}$$

$$C_{\alpha\beta}{}^\gamma = +\tfrac{1}{4}(\gamma_{de})_{(\alpha}{}^\gamma C_{\beta)}{}^{de} \ , \qquad C_\alpha{}^{bc} = -C_\alpha{}^{cb} \ , \tag{2.3c}$$

$$C_{ab}{}^\gamma = +\tfrac{i}{144}(\gamma_b{}^{defg}F_{defg} + 8\gamma^{def}F_{bdef})_\alpha{}^\gamma - \tfrac{1}{8}(\gamma^{cd})_\alpha{}^\gamma(2C_{bcd} - C_{cdb}) \ . \tag{2.3d}$$

We can confirm the satisfaction of Bianchi identities, whose details are skipped here, as has been explained in [19].

## 3. Invariant Lagrangian

Since our lagrangian in superspace is simple, we first present it, accompanied by notational explanations. Our total action given as a superspace integral $I \equiv \int d^{11}x\, d^{32}\theta\, \mathcal{L}$ is

$$I = \int d^{11}x\, d^{32}\theta\, E^{-1} \mathcal{E}^{A_1 \cdots A_{11}} \Big[ + \mathcal{G}_{A_{11} A_{10}} \mathcal{F}_{A_9 \cdots A_5} \mathcal{A}_{A_4 \cdots A_1}$$
$$+ \mathcal{H}_{A_{11} \cdots A_6} \mathcal{T}_{A_5 A_4 A_3}{}^B \mathcal{S}_{A_2 A_1}{}^C \eta_{CB} \Big] \ . \tag{3.1}$$

As the integration measure indicates, our lagrangian is in the superspace for 11D space-time with 10 bosonic coordinates $x^m$ and 32 Majorana coordinates $\theta^\alpha$ [15]. Our lagrangian is a generalization of $BF$ theory in superspace, in terms of the two superfield strengths $\mathcal{F}_{ABCDE}$, $\mathcal{T}_{ABC}{}^D$, $\mathcal{G}_{AB}$ and $\mathcal{H}_{A_1 \cdots A_6}$ defined by

$$\mathcal{F}_{ABCDE} \equiv \tfrac{1}{4!} E_{[A} \mathcal{A}_{BCDE)} - \tfrac{1}{3! \cdot 2} C_{[AB|}{}^F \mathcal{A}_{F|CDE)} \ . \tag{3.2a}$$

$$\mathcal{T}_{ABC}{}^D \equiv \tfrac{1}{2} E_{[A} \mathcal{S}_{BC)}{}^D - \tfrac{1}{2} C_{[AB|}{}^E \mathcal{S}_{E|C)}{}^D \ . \tag{3.2b}$$

$$\mathcal{G}_{AB} \equiv E_{[A} \mathcal{B}_{B)} - C_{AB}{}^C \mathcal{B}_C \ , \tag{3.2c}$$

$$\mathcal{H}_{A_1 \cdots A_6} \equiv \tfrac{1}{5!} E_{[A_1} \mathcal{C}_{A_2 \cdots A_6)} - \tfrac{1}{4! \cdot 2} C_{[A_1 A_2|}{}^D \mathcal{C}_{D|A_3 \cdots A_6)} \ . \tag{3.2d}$$

In other words, the potential superfields $\mathcal{A}_{ABCD}$, $\mathcal{S}_{AB}{}^C$, $\mathcal{B}_A$ and $\mathcal{C}_{A_1 \cdots A_5}$ respectively have the superfield strengths $\mathcal{F}_{ABCDE}$, $\mathcal{T}_{ABC}{}^D$, $\mathcal{G}_{AB}$ and $\mathcal{H}_{A_1 \cdots A_6}$. The $E$ is the superdeterminant of the vielbein $E_A{}^M$: $E \equiv \text{sdet}\,(E_A{}^M)$ following the notation in superspace [18], and this causes the inverse power of $E^{-1}$ in the integrand in (3.1). The reason we need the products of two pairs of superfield strengths $\mathcal{G}, \mathcal{F}$ and $\mathcal{H}, \mathcal{T}$ will be clarified shortly.

One of the most important quantities used in our action (3.1) is the invariant constant tensors $\mathcal{E}^{A_1 \cdots A_{11}}$ and $\eta_{AB}$ in 11D superspace. In particular, when all the indices are bosonic, the former should be totally antisymmetric, while the latter is totally symmetric, in order our lagrangian (3.1) to be non-vanishing. More explicitly they are defined by[3]

$$\mathcal{E}^{A_1 \cdots A_{11}} \equiv \tfrac{1}{11!} \mathcal{E}^{[A_1 \cdots A_{11})}$$

$$\equiv \begin{cases} \mathcal{E}^{a_1 \cdots a_{11}} \equiv \epsilon^{a_1 \cdots a_{11}} \ , \\ \mathcal{E}^{a_1 \cdots a_9 \beta\gamma} \equiv \epsilon^{a_1 \cdots a_9 bc}(\gamma_{bc})^{\beta\gamma} \ , \\ \mathcal{E}^{a_1 \cdots a_7 \beta_1 \cdots \beta_4} \equiv \epsilon^{a_1 \cdots a_7 b_1 \cdots b_4}(\gamma_{b_1 b_2})^{(\beta_1 \beta_2|}(\gamma_{b_3 b_4})^{|\beta_3 \beta_4)} \ , \\ \mathcal{E}^{a_1 \cdots a_5 \beta_1 \cdots \beta_6} \equiv \epsilon^{a_1 \cdots a_5 b_1 \cdots b_6}(\gamma_{b_1 b_2})^{(\beta_1 \beta_2|}(\gamma_{b_3 b_4})^{|\beta_3 \beta_4|}(\gamma_{b_5 b_6})^{|\beta_5 \beta_6)} \ , \\ \mathcal{E}^{a_1 \cdots a_3 \beta_1 \cdots \beta_8} \equiv \epsilon^{a_1 a_2 a_3 b_1 \cdots b_8}(\gamma_{b_1 b_2})^{(\beta_1 \beta_2|}(\gamma_{b_3 b_4})^{|\beta_3 \beta_4|}(\gamma_{b_5 b_6})^{|\beta_5 \beta_6|}(\gamma_{b_7 b_8})^{|\beta_7 \beta_8)} \ , \\ \mathcal{E}^{a \beta_1 \cdots \beta_{10}} \equiv \epsilon^{a b_1 \cdots b_{10}}(\gamma_{b_1 b_2})^{(\beta_1 \beta_2|}(\gamma_{b_3 b_4})^{|\beta_3 \beta_4|}(\gamma_{b_5 b_6})^{|\beta_5 \beta_6|}(\gamma_{b_7 b_8})^{|\beta_7 \beta_8|}(\gamma_{b_9 b_{10}})^{|\beta_9 \beta_{10})} \ , \\ 0 \quad \text{(otherwise)} \ , \end{cases} \quad (3.3\text{a})$$

$$\eta_{AB} \equiv \tfrac{1}{2}\eta_{(AB]} \equiv \begin{cases} \eta_{ab} \equiv \text{diag.}\,(+,-,\cdots,-) \ , \\ \eta_{\alpha\beta} \equiv C_{\alpha\beta} \ , \\ 0 \quad \text{(otherwise)} \ , \end{cases} \quad (3.3\text{b})$$

where $\eta_{ab}$ is the usual $11 \times 11$ symmetric 11D metric tensor, while $C_{\alpha\beta}$ is the usual $32 \times 32$ antisymmetric charge-conjugation matrix used as a metric for spinors. The above expression for $\mathcal{E}^{A_1 \cdots A_{11}}$ is from 10D supergravity formulation, in particular, what is called 'dual formulation' of Type I supergravity [21]. By comparing the usual version and the dual version of Type I supergravity in superspace [21], we find that the natural generalization of the usual 10D $\epsilon$-tensor $\epsilon^{a_1 \cdots a_{10}}$ is given by $\mathcal{E}^{a_1 \cdots a_8 \beta\gamma} \equiv \epsilon^{a_1 \cdots a_8 bc}(\gamma_{bc})^{\beta\gamma}$, $\mathcal{E}^{a_1 \cdots a_6 \beta_1 \cdots \beta_4} \equiv \epsilon^{a_1 \cdots a_6 b_1 \cdots b_4}(\gamma_{b_1 b_2})^{(\beta_1 \beta_2|}(\gamma_{b_3 b_4})^{|\beta_3 \beta_4)}$, and so forth [21]. In other words, each pair of bosonic indices in the purely bosonic $\epsilon$-tensor is replaced by a pair of fermionic indices by contractions with the $\gamma$-matrix $(\gamma_{bc})^{\beta\gamma}$. There may be some freedom for normalizations for the r.h.s. of (3.3a), but this will not matter in our formulation. More importantly, all of the components of $\mathcal{E}^{A_1 \cdots A_{11}}$ are constant.

At first glance, the introduction of the $\mathcal{E}$-tensor (3.3a) into superspace looks unconventional. However, we point out that this $\mathcal{E}$-tensor has many analogs in dimensions other than the above-mentioned 10D case. For example in supersymmetric Chern-Simons theories in 3D [22], it is known that $N=1$ Yang-Mills-Chern-Simons theory has an analogous usage of $\gamma$-matrices in superspace action as if it were a superspace $\mathcal{E}$-tensor. Note also that our invariant tensor $\mathcal{E}^{A_1 \cdots A_{11}}$ has properties slightly different from the purely bosonic case. First of all, it is *not* 'maximally' antisymmetric in $11 + 32 = 43$-dimensional superspace. This is because in superspace there is no 'maximal rank $\mathcal{E}$-tensor' used in the same way as for purely bosonic coordinates. Note that $_{[A_1 \cdots A_r)} \neq 0$ for the totally (anti)symmetric indices even for $r \geq 12$ in 11D, because $_{[A_1 \cdots A_r)}$ is actually 'symmetrization' for fermionic coordinates, and therefore $_{[A_1 \cdots A_r)}$ does not vanish for $r \geq 12$. This is also reflected in the definition of super-determinant $E \equiv \text{sdet}(E_A{}^M) \equiv \exp[\text{str} \ln(E_A{}^M)]$ with *no* reference to the maximal rank (anti)symmetric $\mathcal{E}$-tensor. We have to keep this in mind, when performing actual computation in superspace.

---

[3]We mention that similar $\epsilon$tensors for superfield actions of Chern-Simons type were proposed in [20].

The fundamental superfields in our theory are the potential superfields $\mathcal{A}_{ABCD}$, $\mathcal{S}_{AB}{}^C$, $\mathcal{B}_A$, $\mathcal{C}_{A_1 \cdots A_5}$, and the vielbein $E_A{}^M$. Since $\mathcal{F}_{ABCDE}$, $\mathcal{T}_{ABC}{}^D$, $\mathcal{G}_{A_1 \cdots A_7}$ and $\mathcal{H}_{A_1 \cdots A_9}$ are all superfield strengths by definition, they should obey their own Bianchi identities. In fact, it is not difficult to show that they satisfy the following Bianchi identities

$$\tfrac{1}{5!} E_{[A_1} \mathcal{F}_{A_2 \cdots A_6)} - \tfrac{1}{4! \cdot 2} C_{[A_1 A_2|}{}^B \mathcal{F}_{B|A_3 \cdots A_6)} \equiv 0 \ , \tag{3.4a}$$

$$\tfrac{1}{3!} E_{[A} \mathcal{T}_{BCD)}{}^E - \tfrac{1}{2! \cdot 2} C_{[AB|}{}^F \mathcal{T}_{F|CD)}{}^E \equiv 0 \ , \tag{3.4b}$$

$$\tfrac{1}{2} E_{[A} \mathcal{G}_{BC)} - \tfrac{1}{2} C_{[AB|}{}^D \mathcal{G}_{D|C)} \equiv 0 \ , \tag{3.4c}$$

$$\tfrac{1}{6!} E_{[A_1} \mathcal{H}_{A_2 \cdots A_7)} - \tfrac{1}{5! \cdot 2} C_{[A_1 A_2|}{}^B \mathcal{H}_{B|A_3 \cdots A_7)} \equiv 0 \ . \tag{3.4d}$$

The satisfaction of the Bianchi identity (3.4b) also justifies the definition (3.2b) of $\mathcal{T}_{ABC}{}^D$ which has a peculiar superscript index $D$ like a free index that stands alone, not interfering with all other subscript indices $ABC$, which is rather new as superspace formulation. In other words, this index $D$ behaves like the familiar adjoint representation index in Yang-Mills theory.

Our action (3.1) has also gauge invariances, under the transformations

$$\delta_\Lambda \mathcal{B}_A \equiv E_A \Lambda \ , \tag{3.5a}$$

$$\delta_\Lambda \mathcal{C}_{A_1 \cdots A_5} \equiv \tfrac{1}{24} E_{[A_1} \Lambda_{A_2 \cdots A_5)} - \tfrac{1}{12} C_{[A_1 A_2|}{}^B \Lambda_{B|A_3 A_4 A_5)} \ . \tag{3.5b}$$

Here $\Lambda$'s with different number of indices are distinct from each other as arbitrary space-time dependent parameters. However, as is seen from the pure potential superfields $\mathcal{A}_{ABCD}$ and $\mathcal{S}_{AB}{}^C$ present with no derivatives, our action (3.1) is *not* invariant ($\delta'_\Lambda I \neq 0$) under the gauge transformations

$$\delta'_\Lambda \mathcal{A}_{ABCD} = \tfrac{1}{6} E_{[A} \Lambda_{BCD)} - \tfrac{1}{4} C_{[AB|}{}^E \Lambda_{E|CD)} \ , \tag{3.6a}$$

$$\delta'_\Lambda \mathcal{S}_{AB}{}^C \equiv E_{[A} \Lambda_{B)}{}^C - C_{AB}{}^D \Lambda_D{}^C \ , \tag{3.6b}$$

for the potential superfields $\mathcal{A}_{A_1 \cdots A_5}$ and $\mathcal{S}_{AB}{}^C$. We can confirm that $\delta'_\Lambda I \neq 0$ more rigorously by direct computation, with the aid of the Bianchi identities (3.4) and the definitions (3.2).

It is worthwhile to mention that our action (3.1) can be rewritten after appropriate partial integrations, as

$$I = \int d^{11}x \, d^{32}\theta \, E^{-1} \mathcal{E}^{A_1 \cdots A_{11}} \Big[ + \tfrac{1}{12!} C_{[A_{11} A_{10}|}{}^B \mathcal{B}_{|B|} \mathcal{F}_{|A_9 \cdots A_5|} \mathcal{A}_{|A_4 \cdots A_1)}$$

$$+ \tfrac{3}{12!} \mathcal{C}_{[A_{11}A_{10}|}{}^B \mathcal{C}_{|BA_9\cdots A_6|} \mathcal{T}_{|A_5 A_4 A_3|}{}^D \mathcal{S}_{|A_2 A_1)}{}^C \eta_{CD} \Big] ~~, \tag{3.7}$$

where all the 12 indices are totally antisymmetrized in each term. For the reason already mentioned, such a total antisymmetrization does not lead to a vanishing result, due to the special feature of our $\mathcal{E}$-tensor used here, as well as common feature of superspace [18][15]. Compared with (3.7), the previous expression (3.1) of our action $I$ looks more like a generalized $BF$ theory [1], in the sense that each term is a product of a potential superfield and products of superfield strengths. However, the alternative expression (3.7) is much more topological in the sense that the lagrangian in (3.7) changes as a total divergence under the gauge transformation (3.5), much like the case of $BF$ theories or Chern-Simons theories [1]. This feature is not manifest in (3.1), because superfield strengths are manifestly invariant under (3.5). Note that the index $B$ in the second line in (3.7) is under the total (anti)symmetrizations of 12 indices $[A_{11}A_{10}BA_9A_8\cdots A_1)$. For the reason already mentioned, such an expression with 12 total (anti)symmetrization does not necessarily vanish in superspace.

Some readers may wonder, if this type of action starting with the trilinear terms such as (3.1) really makes sense. Because, *e.g.*, for quantizations it is more convenient to start with the bilinear terms in the lagrangian, instead of trilinear terms. This question is answered by the other Chern-Simons theories in higher-dimensions [7], in which the quantizations can be performed by appropriate expansions of relevant fields around their non-vanishing v.e.v.'s, so that the fluctuations for quantized fields will start at the bilinear order [7]. In other words, our action (3.1) starting with trilinear terms suggests some transitions between vacua, as different phases of this potentially underlying master theory. For example, we can try the v.e.v.'s $\langle \mathcal{G}_{\hat{a}\hat{b}} \rangle = \epsilon_{\hat{a}\hat{b}} \neq 0$, $\langle \mathcal{G}_{\hat{a}\bar{b}} \rangle = 0$, $\langle \mathcal{G}_{\bar{a}\bar{b}} \rangle = 0$ for the dimensional reduction from 11D into 9D. Here the indices $\hat{a}, \hat{b}, \cdots$ are for the extra 2D in this dimensional reduction, while $\bar{a}, \bar{b}, \cdots$ are for the 9D indices. By this reduction, the purely bosonic part in $\mathcal{GFA}$-term in the original action (3.1) yields a bilinear Chern-Simons term in 9D: $\epsilon^{\bar{a}_1\cdots\bar{a}_9} \mathcal{F}_{\bar{a}_9\cdots\bar{a}_5} \mathcal{A}_{\bar{a}_4\cdots\bar{a}_1}$. In a similar fashion, we can give nontrivial v.e.v.'s to other field strengths or covariant superfields that leads us to many other interesting actions starting with bilinear terms.[4] We will give an illustrative example of such a dimensional reduction in the next section.

The equivalence between the two expressions (3.1) and (3.7) can be confirmed by direct computation, including appropriate partial integrations. A useful relationship to be used is

$$\partial_M (E^{-1} E_A{}^M) = C_{AB}{}^B \equiv E^{-1} C_A ~~. \tag{3.8}$$

Even though we did not write the Grassmann parities explicitly, they are to be understood in the standard way. Namely, when the nearest neighbor pair of two indices is contracted between northwest and southeast, there is no cost of sign, while that between northeast and southwest costs an additional sign. (We call this 'nest form contraction'.) To be more specific, the Grassmann parities in (3.8) can be explicitly written as

$$(-1)^{M(A+M)}\partial_M(E^{-1}E_A{}^M) = (-1)^B E^{-1} C_{AB}{}^B \equiv E^{-1}C_A \quad . \tag{3.9}$$

---

[4]We can give such non-vanishing v.e.v.'s, as long as all the superfield equations (3.12) are satisfied.

In other words, whenever we have expression like (3.8), we consider all the sign changes to put all the indices in the 'nest form', *i.e.*, all the contracted indices are in the nearest pairs between northwest and southwest positions. As long as we keep this 'nest form' rule in mind, there will arise no ambiguities in signatures, as (3.9) is uniquely obtained from (3.8). In deriving (3.7) from (3.1), we also need the identities

$$\mathcal{F}_{[A_1\cdots A_5|}\mathcal{F}_{|A_6\cdots A_{10})} \equiv 0 \quad , \qquad \mathcal{T}_{[A_1A_2A_4|}{}^E \mathcal{T}_{|A_4A_5A_6)}{}^F \eta_{FE} \equiv 0 \quad . \tag{3.10}$$

In the second equation, appropriate Grassmann parities are to be understood for the exchange of the indices $E$ and three out of $A_1$, $A_2$, $\cdots$, $A_6$, even though they are not explicitly written. For example, the last equation of (3.10) implies

$$\mathcal{T}_{[A_1A_2A_3|}{}^E \mathcal{T}_{|A_4A_5A_6)}{}^F \eta_{FE} \equiv \sum_{\substack{6!\text{ permut. of}\\A_1\cdots A_6}} (-1)^{E(A_4+A_5+A_6)} \mathcal{T}_{A_1A_2A_3}{}^E \mathcal{T}_{A_4A_5A_6}{}^F \eta_{FE}$$

$$\equiv +(-1)^{E(A_4+A_5+A_6)} \mathcal{T}_{A_1A_2A_3}{}^E \mathcal{T}_{A_4A_5A_6}{}^F \eta_{FE}$$
$$+ (-1)^{E(A_5+A_6+A_1)+A_1(A_2+A_3+A_4+A_5+A_6)} \mathcal{T}_{A_2A_3A_4}{}^E \mathcal{T}_{A_5A_6A_1}{}^F \eta_{FE}$$
$$+ \left[\,(6!-2)\text{ more terms}\,\right] \quad . \tag{3.11}$$

In this paper, we do not write Grassmann parities explicitly, in order to save considerable amount of space, but they are always to be understood as in other superspace formulations [18][15].

We now consider the superfield equations from our superspace action in (3.1). Our fundamental superfields are $\mathcal{B}_{A_1\cdots A_4}$, $\mathcal{S}_{AB}{}^C$, $\mathcal{B}_A$, $\mathcal{C}_{A_1\cdots A_5}$ and $E_A{}^M$ with no additional constraints. Therefore, their superfield equations are obtained as the usual superspace Euler derivatives [18]:

$$\frac{\delta \mathcal{L}}{\delta \mathcal{A}_{A_1\cdots A_4}} = +\tfrac{5}{9!\cdot 7!\cdot 2} E^{-1} \mathcal{E}^{A_1\cdots A_4 B_1\cdots B_7} \mathcal{G}_{[B_7B_6|}{}^C{}_{|B_5B_4|}{}^C \mathcal{A}_{|CB_3B_2B_1)}$$
$$- \tfrac{10}{9!\cdot 4!} E^{-1} \mathcal{E}^{[A_1A_2A_3|B_1\cdots B_8} \mathcal{G}_{B_8B_7} C_{B_6B_5}{}^{|A_4)} \mathcal{A}_{B_4\cdots B_1} \doteq 0 \quad , \tag{3.12a}$$

$$\frac{\delta\mathcal{L}}{\delta\mathcal{S}_{AB}{}^C} = +\tfrac{3}{9!\cdot 5!\cdot 2} E^{-1} \mathcal{E}^{ABD_3\cdots D_{11}} \mathcal{H}_{[D_{11}\cdots D_6|}{}^C{}_{|D_5D_4|}{}^E \mathcal{S}_{|ED_3)}{}^F \eta_{FC}$$
$$+ \tfrac{3}{5!} E^{-1} \mathcal{E}^{[A|E_1\cdots E_{10}} \mathcal{H}_{E_{10}\cdots E_5} C_{E_4E_3}{}^{|B)} \mathcal{S}_{E_2E_1}{}^F \eta_{FC} \doteq 0 \quad , \tag{3.12b}$$

$$\frac{\delta\mathcal{L}}{\delta\mathcal{B}_A} = \tfrac{1}{12} E^{-1} \mathcal{E}^{B_1\cdots B_{11}} C_{B_{11}B_{10}}{}^A \mathcal{F}_{B_9\cdots B_5} \mathcal{A}_{B_4\cdots B_1}$$
$$- \tfrac{11}{12!} E^{-1} \mathcal{E}^{AB_{10}\cdots B_1} C_{[B_{10}B_9|}{}^C \mathcal{F}_{|CB_8\cdots B_5|} \mathcal{A}_{|B_4\cdots B_1)} \doteq 0 \quad , \tag{3.12c}$$

$$\frac{\delta \mathcal{L}}{\delta \mathcal{C}_{A_1 \cdots A_5}} = + \frac{1}{6! \cdot 4} E^{-1} \mathcal{E}^{A_1 \cdots A_5 B_1 \cdots B_6} \mathcal{C}_{[B_6 B_5|}{}^C \mathcal{T}_{|CB_4 B_3|}{}^D \mathcal{S}_{|B_2 B_1)}{}^E \eta_{ED}$$
$$- \frac{1}{4! \cdot 4} E^{-1} \mathcal{E}^{[A_1 \cdots A_4|B_1 \cdots B_7} \mathcal{C}_{B_7 B_6|}{}^{|A_5)} \mathcal{T}_{B_5 B_4 B_3}{}^D \mathcal{S}_{B_2 B_1}{}^C \eta_{CD} \doteq 0 ~, \quad (3.12\text{d})$$

$$E_B{}^M \frac{\delta \mathcal{L}}{\delta E_A{}^M} = - \delta_B{}^A \mathcal{L} + 2 E^{-1} E_C X_B{}^{CA} - 2 E^{-1} C_C X_B{}^{CA}$$
$$+ 2 E^{-1} C_{BC}{}^D X_D{}^{CA} - E^{-1} C_{DC}{}^A X_B{}^{CD} \doteq 0 ~, \quad (3.12\text{e})$$

where

$$X_A{}^{BC} \equiv - \mathcal{E}^{BCD_1 \cdots D_9} \mathcal{B}_A \mathcal{F}_{D_9 \cdots D_5} \mathcal{A}_{D_4 \cdots D_1} - 10 \mathcal{E}^{BCD_1 \cdots D_9} \mathcal{G}_{D_9 D_8} \mathcal{A}_{AD_7 D_6 D_5} \mathcal{A}_{D_4 \cdots D_1}$$
$$- 15 \mathcal{E}^{BCD_1 \cdots D_9} \mathcal{C}_{AD_9 \cdots D_6} \mathcal{T}_{D_5 D_4 D_3}{}^E \mathcal{S}_{D_2 D_1}{}^F \eta_{FE}$$
$$- 3 \mathcal{E}^{BCD_1 \cdots D_9} \mathcal{H}_{D_9 \cdots D_4} \mathcal{S}_{AD_3}{}^E \mathcal{S}_{D_2 D_1}{}^F \eta_{FE} ~. \quad (3.13)$$

The symbol $\doteq$ implies an equality that holds only by the use of superfield equation, but not an identity. In all of the equations in (3.12), appropriate Grassmann parities mentioned with (3.11) are to be understood, even though they are implicit. Note that the $C$-index in (3.12a), (3.12c), (3.12d), or the $E$-index in (3.12b) are inside of the total (anti)symmetrizations. It is no wonder that the bare potential superfields $\mathcal{A}_{ABCD}$ and $\mathcal{S}_{AB}{}^C$ with no derivatives appear in these superfield equations, considering the fact that our action does not have gauge invariance under (3.6). We will come back to this point shortly.

We are now ready to discuss how these superfield equations can embed our teleparallel 11D superspace supergravity [19], characterized by the Bianchi identities (2.1). First of all, note that the set of solutions

$$\mathcal{T}_{ABC}{}^D \doteq 0 ~, \quad \mathcal{F}_{ABCDE} \doteq 0 ~, \quad (3.14\text{a})$$
$$\mathcal{B}_A \doteq 0 ~, \quad \mathcal{C}_{A_1 \cdots A_5} \doteq 0 ~, \quad (3.14\text{b})$$

trivially satisfies our superfield equations (3.12) including (3.12e) for the vielbein, because of $X_A{}^{BC} \doteq 0$ under (3.14). At first glance, this set of solutions looks trivial, leading to no physical content. However, the consideration of the identifications

$$\mathcal{S}_{AB}{}^C \doteq C_{AB}{}^C ~, \quad \mathcal{A}_{ABCD} \doteq F_{ABCD} ~, \quad (3.15)$$

reveals non-trivial nature of our system, because all the components in the Bianchi identities in 11D superspace (2.1) are now satisfied, combined with (3.14), (3.15) and (3.2):

$$\mathcal{T}_{ABC}{}^D \doteq \tfrac{1}{2} E_{[A} C_{BC)}{}^D - \tfrac{1}{2} C_{[AB|}{}^E C_{E|C)}{}^D \equiv 0 ~, \quad (3.16\text{a})$$
$$\mathcal{F}_{ABCDE} \doteq \tfrac{1}{24} E_{[A} F_{BCDE)} - \tfrac{1}{12} C_{[AB|}{}^F F_{F|CDE)} \equiv 0 ~. \quad (3.16\text{b})$$

In other words, all the teleparallel superspace Bianchi identities [19] for 11D supergravity (2.1) are embedded into the superfield equations from our action (3.1), as the simplest but nontrivial solutions.

One important ingredient here is that there can be other nontrivial solutions to our superfield equations in (3.12) other than (3.14), because there are many indices contracted in (3.12), but not necessarily all the components of each of the superfield strengths $\mathcal{T}_{ABC}{}^D$ and $\mathcal{F}_{ABCDE}$ are zero. This also indicates that our system of superspace $BF$ theory can accommodate more theories than the ordinary 11D supergravity [5][15], suggesting strongly that our theory is one of the most natural generalizations of 11D supergravity [5][15], as a good low energy theory for M-theory [3]. As we have also mentioned, the feature of our action starting with the trilinear terms also suggests some non-trivial vacuum with non-vanishing v.e.v.'s of the fundamental superfields, like other Chern-Simons theories in higher dimensions [7].

We now address the usage of teleparallel superspace in 11D [19]. The most important advantage of teleparallelism [19] is that the supertorsion Bianchi identity for teleparallel superspace is much more simplified than the conventional one with the local Lorentz connection $\phi_{Ab}{}^c$. For example, if we had the supercurvature term proportional to $R_{[AB|c}{}^d(\mathcal{M}_d{}^c)_{|C)}{}^D$ in the Bianchi identity (2.1a), then it would look unnatural to embed such a term into another superfield strength like $\mathcal{T}_{ABC}{}^D$ in (3.2b). Or more importantly, the usual expression of $\phi_{Ab}{}^c$ in terms of the vielbein $via$ anholonomy coefficients $C_{ab}{}^c$ came out, only $after$ solving the Bianchi identity $\nabla_\alpha T_{bc}{}^d + \cdots \equiv 0$ at mass dimension $d = 3/2$. Therefore, it is more natural to work directly on the teleparallel superspace [19] from the outset, where there is no worry of solving the Bianchi identity at $d = 3/2$. From these viewpoints, we consider teleparallel superspace [19] is the most natural choice for superspace formulation for our purpose.

It is worthwhile to mention that a somewhat analogous trial of using a third-rank antisymmetric potential field in 11D supergravity [5] interpreted as a torsion tensor was presented as early as in 1984 [23]. There are differences as well as similarities between our formulation and [23]. The similarity is that we are identifying superfield strengths such as $F_{ABCD}$ with the superpotential $\mathcal{A}_{ABCD}$, analogously to the potential $A_{mnr}$ embedded into the torsion $T_{mn}{}^r$ which is also a field strength [23]. One of the differences is that our formulation is not just a re-writing of the conventional 11D supergravity as in [23], but it presents a more topological generalized $BF$-theory as an underlying theory for 11D supergravity [5][15] itself. Additionally, we are dealing with superspace instead of components formulation [23].

We now discuss the previously-mentioned gauge $non$-invariance of our action (3.1) under (3.6). This is to guarantee that our superpotential $A_{ABC}$ after the embedding (3.15) is $not$ gauged away. To see this, let us consider a much simpler 'toy' action

$$I' = \int d^{11}x \, d^{32}\theta \, E^{-1} \mathcal{E}^{A_1 \cdots A_{11}} \mathcal{B}_{A_{11} \cdots A_6} \mathcal{F}_{A_5 \cdots A_1} \quad . \tag{3.17}$$

This action has a manifest gauge invariance under (3.6), and it looks much simpler than our action $I$ in (3.1). What is wrong with this action $I'$ is that the gauge symmetry (3.6) will gauge away the embedded potential superfields $A_{ABC}$ completely, if we choose

$$\Lambda_{ABC} = -A_{ABC} \ . \tag{3.18}$$

In other words, for a system with an action with the gauge symmetry (3.6), such embedding as (3.15) becomes just a gauge degree of freedom, leading to no physical content. This is a crucial point for our formulation, because we are embedding the superfield strength $F_{ABCD}$ into the potential superfield $\mathcal{A}_{ABCD}$ whose superfield strength $\mathcal{F}_{ABCDE}$ is one hierarchy higher, which can easily lead to a trivial result, by 'cohomological nilpotency'. Such obstructions may be the main cause of the delay for the development of superspace formulations of $BF$ theory [1] up until now. This also tells us why we needed the products of two pairs of superfield strengths $\mathcal{G}$, $\mathcal{F}$ and $\mathcal{H}$, $\mathcal{T}$ in (3.1) as a generalization of $BF$ theory. It further tells us how difficult it is to 'embed' such a simple looking set of Bianchi identities as (2.1) into some superfield equations obtained under an action principle.

This sort of gauge-non-invariance $\delta'_\Lambda I \neq 0$ under (3.6) seems rather awkward at a first glance, because there do not seem to be so many other examples in physics. However, we emphasize that this is what is happening rather often in supergravity theories. First of all, we have to remind ourselves that our system is not locally Lorentz covariant from the outset. This can be explicitly confirmed by the variation of the vielbein:

$$\delta_\text{L} E_A{}^M = L_A{}^B E_B{}^M \ , \quad \delta_\text{L} E_M{}^A = -E_M{}^B L_B{}^A \ , \tag{3.19}$$

with the infinitesimal local Lorentz parameter $L^{AB} = L^{AB}(Z) = -(-1)^{AB} L^{BA} \equiv \left( L^{ab}, L^{\alpha\beta} \equiv (1/4)(\gamma_c{}^d)^{\alpha\beta} L_d{}^c \right)$. Accordingly, the anholonomy coefficients $C_{AB}{}^C$ acquires an inhomogeneous term at the end:

$$\delta_\text{L} C_{AB}{}^C = L_{[A|}{}^D C_{D|B)}{}^C - C_{AB}{}^D L_D{}^C + E_{[A} L_{B)}{}^C \ , \tag{3.20}$$

and therefore our action (3.7) lacks local Lorentz invariance under (3.19). Analogously to this local Lorentz symmetry, the local symmetry for our solution $A_{ABC}$ in (3.15):

$$\delta''_\Lambda A_{ABC} = \tfrac{1}{2} E_{[A} \Lambda''_{BC)} - \tfrac{1}{2} C_{[AB|}{}^D \Lambda''_{D|C)} \ , \tag{3.21}$$

is not manifest in our original action, while this symmetry is realized within a particular set of solutions at the superfield equation level. This phenomenon is not new, but has been well-known in supergravity theories, such as local $SU(8)$ symmetry in $N=8$ supergravity in 4D [5] related to U-duality. To put it differently, our starting action does not have to possess all the local symmetries needed at the end, but those symmetries such as (3.19) or (3.21) are realized only among their particular solutions. In fact, once we have obtained the 11D supergravity as a solution to the superfield equations (3.12), the local symmetry (3.21) for $A_{ABC}$ is recovered as in usual 11D supergravity [15]. Once we have accepted this viewpoint, it is becomes natural that our system lacks certain symmetries or invariances such as (3.6a) at the level of action or lagrangian.

## 4.  Concluding Remarks

In this paper, we have presented a simple lagrangian formulation of $BF$ theory [1] in superspace [18] that can accommodate all the necessary Bianchi identities for teleparallel 11D superspace supergravity [19]. We have two expressions (3.1) and (3.7) for our action $I$. The latter is much more 'topological' than its alternative expression (3.1), in the sense that the lagrangian changes as a total divergence in superspace under gauge transformations. We have seen that the 11D superspace Bianchi identities are realized as one of the simplest but nontrivial solutions to our superfield equations from our superspace lagrangian. This also suggests that the superfield equations for our lagrangian allow more solutions than those for 11D supergravity [15], and therefore, our topological theory can be one of the most natural generalization of the conventional 11D supergravity [5][15]. Our action (3.1) starts with the trilinear order terms, instead of the usual bilinear ones, suggesting the transition of vacuum from the original one where superfields are expanded. This feature is also common to other Chern-Simons theories in higher dimensions [7].

It has been generally expected that M-theory [3] will reveal itself as a topological or geometrical theory, such as Chern-Simons theory [6][7][8], $BF$ theory [1], or topological theory [10] like Poincaré gravity can be realized as a more geometrical conformal gravity theory, when the dimensionful gravitational coupling becomes negligible, or at higher energy, or at quantum level. In this paper, we have presented a generalized $BF$ theory as a topological/geometrical generalization of 11D supergravity [15], relying on formulation in teleparallel superspace [19]. Our action (3.1) starting with trilinear terms indicates possible nontrivial v.e.v.'s yielding bilinear terms. This suggests the existence of many different phases of our theory with nontrivial vacuum structures.

To our knowledge, we stress that such a formulation of 'generalized' topological $BF$ theory [1] in 11D superspace [15] has never been presented. There seem to have been three main obstructions against such formulations in superspace in the past. First, the supercurvature term in the $T$-Bianchi identity [18] prohibits the simple embedding of the supertorsion tensor $T_{AB}{}^C$ into a higher-rank tensor. We have overcome this obstruction by getting rid of the supercurvature term in (2.1a), adopting teleparallel superspace [19]. Second, superspace formulation [18] seemed to lack the appropriate invariant constant $\mathcal{E}$-tensor for fermionic coordinates, as a simple analog of $\epsilon$-tensor for topological gauge theories only with the bosonic coordinates. We have overcome this obstruction by the use of $\mathcal{E}$-tensor (3.3a) acquired from experience in 3D Chern-Simons theory in superspace [22] and the dual Type I supergravity [21], as the most natural higher-dimensional application. Third, as has been mentioned at the end of the last section, the 'cohomological nilpotency' invalidates the idea of such embeddings as (3.15), due to the gauge symmetry that gauge away physical degrees of freedom. This obstruction has been overcome by forbidding the troublesome gauge symmetry (3.6) by the use of Chern-Simons factor as in (3.1). In other words, our system lacks the gauge invariance under such a troublesome gauge symmetry (3.6).

We stress that the usage of Lorentz non-covariant formulation presented in this paper is nothing peculiar or eccentric, but based on the recent developments related M-theory. This is not only motivated by higher dimensional supergravity theories in [24] for F-theory [25] or S-theory [26], but also by other results related to M-theory [3] or superstrings in $D \leq 11$. For example, in ref. [27] a new supergravity formulation with background with Killing isometry has been presented in 11D. Even though the exact terminology 'teleparallelism' is not used in the paper, such a formulation in 11D lacks manifest local Lorentz covariance within 11D, due to the Killing isometry. Another example is found in a recent paper [28] in which some new maximally supersymmetric backgrounds are found for Type IIA superstring. Despite the fact that the phrase 'Lorentz non-covariance' was not used in [28], their result shows the importance of maximally supersymmetric backgrounds which lacks the full 10D local Lorentz covariance. These are just simple examples in this new direction associated with M-theory [3], and the importance of exploring Lorentz non-covariant formulation is now clear. We emphasize that the result in this paper is based on a profound supporting evidence for Lorentz non-covariant teleparallelism in general supergravity/superstring theories, such as recent developments associated with M-theory [3].

In our paper, we have identified supertorsions (anholonomy coefficients) $C_{AB}{}^C$ with the supertensor $\mathcal{S}_{AB}{}^C$. Even though this kind of identification seems unusual or artificial at first sight, we stress that there have been analogous methods since 1980's. One good example can be found in a paper [29] on $N=1$ superspace in 10D. It has been demonstrated in [29] that the consistent embedding of the supertorsion $T_{ab}{}^c$ into a third-rank superfield strength $H_{abc}$ is possible by the relationship $H_{abc} = -(1/2)T_{abc}$. It is to be stressed that our approach has provided a generalization of such an embedding to the case of 11D supergravity [5][15].

It seems that the teleparallelism formulation [19] is universally possible in any space-time dimensions, once the conventional Lorentz covariant formulation in superspace [18] is established for any supergravity [30]. However, in this paper we have taken advantage of 11D teleparallel superspace [19] in which there are only two Bianchi identities needed for supergravity. It is this simplicity of superspace that enabled us to formulate our $BF$ theory that can embed 11D supergravity theory [5][15].

Our system has another noteworthy aspect about local symmetries. We have seen that not only local Lorentz symmetry, but also local symmetry for $A_{ABC}$, that are not present at the action level, can be recovered at the level of superfield equations. We stress that this phenomenon is not new at all, but has been observed in many different contexts in supergravity theories in the past, such as the local $SU(8)$ symmetry for $N=8$ supergravity in 4D [5], that can be realized only among the field equations, but not at the action or lagrangian level.

Our result in this paper has opened a wide avenue for applications of this formulations to other supergravities in diverse dimensions in $1 \leq D \leq 10$ [18][30]. For example, our formulation provides an action principle even for certain supergravity theories, such as Type IIB in 10D, that forbid conventional action formulations. Another interesting study will be on the dimensional reduction of our action into $1 \leq D \leq 10$ that will generate not only all the other known supergravity theories in these lower dimensions [30], but also unknown ones.

Even though our topological superfield theory lagrangian (3.1) looks simple, it is not merely a single term of $BF$ theory, but is an intricate combination of two terms. Neither is it in component language, but it is in terms of superspace language [18] with manifest supersymmetry. It can be understood as a 'generalized' topological $BF$ theory or Chern-Simons theory [1] in the sense that the lagrangian in (3.7) changes as a total divergence under certain gauge transformation. It has space-time supersymmetry instead of 'fake' supersymmetry of supergroups used in Chern-Simons supergravity formulations [6][7][8]. It is not formulated in ordinary superspace [18], but it is more naturally formulated in peculiar teleparallel superspace [19]. We find that our formulation is so simple and elaborate at the same time that this theory might well be the genuine fundamental underlying theory of M-theory [3] as the low energy limit, accommodating the conventional 11D supergravity [5][15].

We are grateful to the referee of this paper who gave important suggestions to improve the original manuscript.